\documentclass[12pt]{article}

\usepackage{amsmath,amsfonts,amssymb}
\usepackage{graphicx}
\usepackage{psfrag}
\usepackage{enumerate}
\usepackage{mathrsfs}

\makeatletter \@addtoreset{equation}{section}

\makeatletter\renewcommand\section{\@startsection {section}{1}{\z@}%
                                   {-3.5ex \@plus -1ex \@minus -.2ex}
                                   {2.3ex \@plus.2ex}%
                                   {\normalfont\large\bfseries}}
\renewcommand\subsection{\@startsection{subsection}{2}{\z@}%
                                     {-3.25ex\@plus -1ex \@minus -.2ex}%
                                     {1.5ex \@plus .2ex}%
                                     {\normalfont\bfseries}}

\newcommand{\be}{\begin{equation}}
\newcommand{\ee}{\end{equation}}
\newcommand{\beq}{\begin{eqnarray}}
\newcommand{\eeq}{\end{eqnarray}}

\def\[{\left [}
\def\]{\right ]}
\def\({\left (}
\def\){\right )}

\def\r2{\sqrt{2}}

\def\Label#1{\label{#1}%
  \smash{\hbox to0pt{\raise1ex\hbox{\tiny[#1]}\hss}}}
\def\Bibitem#1{\bibitem{#1}%
  \smash{\hbox to0pt{\raise1ex\hbox{\tiny[#1]}\hss}}}  
\reversemarginpar
\newcommand{\bbibitem}[1]{\bibitem{#1}\marginpar{\tiny{[#1]}}}

\def\noLabels{\let\Label=\label}
\def\nobbibitem{\let\bbibitem=\bibitem}
\def\noBibitem{\let\Bibitem=\bibitem}

\begin{document}
\noLabels 
\noBibitem 




\begin{center}

{\Large \bf What we don't know about time}

\vspace{3mm}

Vijay Balasubramanian$^{a,}$\footnote{\tt email:vijay@physics.upenn.edu}

\vspace{5mm}

\bigskip\centerline{$^a$\it David Rittenhouse Laboratories}
\smallskip\centerline{\it University of Pennsylvania}
\smallskip\centerline{\it  
 Philadelphia, PA 19104, USA}


\end{center}
\begin{abstract}
\noindent
String theory has transformed our understanding of geometry, topology and spacetime.  Thus, for this special issue of {\it Foundations of Physics} commemorating ``Forty Years of String Theory'', it seems appropriate to step back and ask what we do not understand.   
As I will discuss,  time remains the least understood concept in physical theory.  While we have made significant progress in understanding {\it space}, our understanding of time has not progressed much beyond the level of a century ago when Einstein introduced the idea of space-time as a combined entity.    Thus, I will raise a series of open questions about time, and will review some of the progress that has been made as a roadmap for the future.  
\end{abstract}


\addtocounter{footnote}{-1}

\section{What is time?}

Time is the most mysterious entity in physics.   Space, the better understood cousin of time, is readily conceptualized as a sort of stage.  One  measures distances on this stage, and can move back and forth between locations.  But time is different.  We say that ``time passes",
or admonish children that there is no point  ``crying over spilt milk", or threaten enemies by saying ``you'll be history'', all because we seem not to be able to go back to earlier times.   
Of course, more than a century ago Einstein showed that space and time together form a geometrical entity, spacetime, and also that the reference frames of inertial observers in spacetime are related by hyperbolic rotations that mix space and time.  General relativity goes further and allows complicated warpings that mix time and space in different ways locally.  Still, because information can only propagate at speeds less than $c = 3 \times 10^8 \, {\rm m/s}$, the speed of light,  there is no way of influencing the past or revisiting one's childhood unless the geometry of the universe contains a closed temporal loop (a Closed Timelike Curve, or CTC).  But there are also results that strongly suggest that such universes cannot exist, or more precisely, that any attempt to create one will result in a space-time singularity that drastically alters the geometry \cite{HawkingCTC}.  This leads to a first question: {\it Why is there an arrow of time?}.\footnote{An important approach to the arrow of time involves the second law of thermodynamics.   The idea is that the arrow of time is cosmologically defined by the macroscopic increase of entropy (e.g., see the discussions in \cite{ArrowOfTime}).  This raises the associated question of why the universe starts in a low entropy state.  Since time exists even in empty Minkowksi space, this approach also suggests that the notion of time is inherently connected to the coarse graining of an underlying quantum gravitational configuration space.}

Geometrically, time is different from space because the geometry of spacetime is locally Minkowski (Lorentzian metric signature $(1,3)$), not Euclidean (metric signature $(0,4)$).     If we take lengths in space to square to positive numbers, lengths in time square to negative numbers (i.e., locally $ds^2 = -dt^2 + d\vec{x}^2$).  From a geometrical point of view we could equally well imagine a signature $(2,2)$, with two times, which is more symmetric between space and time.   In the maximally supersymmetric four dimensional Yang-Mills theory, $(2,2)$ signature makes a mysterious appearance in greatly simplifying amplitude calculations via a twistor transform \cite{twistor}.  In the context of string theory with its many extra dimensions one can ask why we seem to have extra spatial dimensions, not temporal dimensions.   Of course, formulating a notion of dynamics with multiple times is confusing \cite{twotimes} since we are used to constructing models by specifying data on a spatial surface and evolving this data with time.   However, geometrically there seems to be no obstacle to having multiple times.   What is more, there are arguments  that certain duality transformation may relate conventional string theories to theories with more than one time \cite{hull}. Thus we have a second question: {\it Why is there only one time?}

The difference between time and space is somehow implicated in the difference between quantum mechanics with its characteristic features of quantum interference and entanglement, and classical statistical physics which lacks these features.  To see this, recall first that the vacuum to vacuum transition amplitude of quantum mechanics is written in path integral language as
\be
Z_q = \int {\cal D}\phi  \, e^{-{i \over \hbar} \int dt \, d\vec{x} \,  {\cal L}(\phi)} \, .
\Label{QFTpath}
\ee
where ${\cal L}$ is the Lagrangian function and the integral sums over all trajectories of field configurations.  The Euclidean continuation $t \to -i \, t$, which also gives $ds^2 \to dt^2 + d\vec{x}^2$ and ${\cal L} \to {\cal L}_E$,  turns this expression into
\be
Z_s = \int {\cal D}\phi \,  e^{- {1 \over \hbar} \int dt \, d\vec{x} \, {\cal L}_E(\phi)} \, .
\Label{QFTstat}
\ee
This is the statistical partition sum of a Euclidean field theory with a temperature $T \sim \hbar$.  In fact, quantum field theories are usually {\it defined} via such an analytic continuation because the path integral in (\ref{QFTpath}) is generally ill-defined.  We can also see the connection between time and quantum mechanics by recalling that for a massive point particle the amplitude for propagation between spacetime points A and B is given by the sum over paths
\be
A = \int_A^B {\cal D}{\cal P} \, e^{-{ m \over \hbar} \int ds}
\Label{QMpath}
\ee
where $ds = \sqrt{-dt^2 + d\vec{x}^2}$ is the line element along the paths ${\cal P}$.  Thus, timelike paths have oscillatory amplitudes and interfere, while spacelike paths are real, and are exponentially suppressed in length.   Formally, if the metric was Euclidean ( $ds = \sqrt{dt^2 + d\vec{x}^2}$) there would be no interference, and (\ref{QMpath}) would describe a statistical partition sum over paths.   The first-quantized path integral of string theory is formulated in a similar manner; so the same connection between time and quantumness is present there too.  This leads to a third question: {\it Is there a connection between the existence of a time, and the quantumness of the universe?}

Related to these ideas is the odd fact that physical quantities seem to be described by analytic functions of space and time.  We are most  familiar with this in the energy-momentum plane, which we can regard as the Fourier transform of spacetime.  For example, it is well known that the structure of scattering amplitudes of physical particles is controlled by the poles and cuts in the complexified energy-momentum plane.  The structure of amplitudes in complexified space-time seems to similarly contain important physical information.  For example, the amplitudes of maximally supersymmetric four-dimensional Yang-Mills theory appear to be controlled by singularities in the $(2,2)$ signature sheet of complexified spacetime \cite{twistor,yangmillsamplitudes}. In another example in the context of the  AdS/CFT correspondence, the field theory dual to spacetime encodes the physics of events behind a black hole horizon in the structure of amplitudes in complexified time \cite{behindthehorizon}.    Furthermore, in a general time-dependent spacetime the Euclidean continuation that is usually used to define the path integral (e.g. (\ref{QFTstat})) may not exist.  That is, because the metric is time dependent, there may not be an analytic continuation to a real, Euclidean geometry -- in general, we must  think about the complexified spacetime and analytic functions on it.\footnote{Likewise, the Lorentzian ``fuzzball'' geometries (see \cite{fuzzballsolns}, and references therein) which have been considered as candidate geometric black hole microstates do not generally have a Euclidean continuation.}  This leads to a fourth question: {\it Could the real, Lorentzian structure of conventional spacetime be simply a convenient way of summarizing analytic information about an underlying complexified geometry?}

Now, a particular striking prediction of General Relativity is that spacetime singularities exist.  These can be timelike (i.e. localized in space), lightlike (i.e. localized on a null curve), or spacelike (i.e. localized in time).    One of the goals of a quantum theory of gravity such as string theory is to resolve such singularities.   Indeed, many singularities that would naively appear in a classical gravitational treatment are understood in string theory to be resolved by one of several mechanisms -- the presence of D-branes (spacetime solitons acting as sources) \cite{Dbranes}, topological transitions \cite{topsing}, the appearance of new light degrees of freedom \cite{conifold} and so on.  Beautiful though these results are, {\it all} of them involve timelike or null singularities.  We have no examples at all of the successful resolution of singularities localized in time such as those that occur at the Big Bang or in the interior of black holes forming from stellar collapse.  Thus we have a fifth question: {\it How can singularities localized in time be resolved in string theory or some other quantum theory of gravity?}

Universes that have a time also have {\it causal structures}, i.e. relationships between spacetime points that encode the possibility of cause and effect between events localized to those points (assuming that signals cannot propagate faster than light).   General Relativity admits many solutions that contain {\it horizons} that separate local observers, thus restricting their causal interactions.  Notable amongst these spacetimes containing horizons are black holes and accelerating geometries such as de Sitter space.   Semiclassical analyses of quantum mechanics in such spacetimes suggest that inertial observers perceive the horizon as having an entropy proportional to area ($S = A/4 G_N \hbar$), and a temperature proportional to the surface gravity at the horizon \cite{BHthermo}.     These phenomena have been subject to intense scrutiny within string theory over the last 15 years.  Indeed,  for large classes of black holes in string theory, it has been possible to show that the entropy of the horizon is equal to the statistical degeneracy of underlying quantum gravitational states \cite{bhentropy}.   However,  in asymptotically flat space this program has only been carried to completion for supersymmetric extremal and near-extremal black holes, while in asymptotically AdS spacetimes there is only a precise account for three dimensional black holes where an infinite dimensional asymptotic symmetry algebra essentially determines the entropy formula completely given the mass and the angular momentum.     There is no statistical explanation to date for the entropy of either black holes formed from stellar collapse, or the entropy of cosmological horizons in accelerating universes.  Neither is there any explanation of why entropy becomes associated to a geometrical construct -- the area of a horizon.    Thus we come to a sixth question:  {\it Why is the area of a horizon, a causal construct, related to entropy, a thermodynamic concept, and can this entropy be given a statistical explanation for general horizons?}

Associated to the presence of a causal horizon is the ``information loss paradox" for black holes, where Hawking radiation apparently leads to non-unitary semiclassical evolution of quantum states  \cite{infoloss}.    The apparent loss of unitarity can be traced ultimately to the causal disconnection of the region behind the horizon, as a result of which events occurring in the interior cannot propagate causally to the exterior.  There have been several attempts within string theory to argue that black hole evolution must be unitary from the perspective of the asymptotic observer (i.e., there is no information paradox, see the review \cite{inforecovery}).    Some of these approaches suggest that the causal disconnection of the black hole interior is a semiclassical artifact of coarse graining over a large space of microstates \cite{fuzzball,noinside,albionevagary,Babeletal}; others suggest that the interior and the exterior of the horizon are ``complementary'' descriptions of the same physics \cite{complementarity}.   Several approaches use the duality between gravity in asymptotically Anti-de Sitter spacetimes and a manifestly unitary local quantum field theory on the spacetime boundary \cite{adscft} to argue that there is simply no room in the full quantum theory for information loss in black holes.   All of these accounts attempt to resolve the information paradox in terms of recovery of information at asymptotic infinity.   However, none of them gives a precise, quantitative account of how unitarity is consistent with the local experiments that a causally disconnected infalling observer can apparently make behind the black hole horizon in the semiclassical theory.   This gives a seventh question: {\it How precisely is physics beyond a black hole horizon encoded in a unitary description of spacetime? }

Much of the progress in resolving the black hole information paradox within string theory has occurred in the context of the AdS/CFT correspondence.  According to this correspondence, string theory in a (d+1)-dimensional, asymptotically Anti-de Sitter (AdS) spacetime is exactly equivalent to a d-dimensional quantum field theory defined on the timelike boundary of such a universe \cite{adscft,adscft2}.    Thus, the radial dimension of AdS spacetime (as well as any additional compact dimensions of the bulk string theory) must be regarded as somehow ``emergent'' from the dynamics of the d-dimensional field theory.    There is an extensive literature investigating how this happens -- in essence, energy scales in the field theory are translated into radial positions in spacetime, so that the renormalization group flow of the field theory becomes geometrized in terms of the radial equations of motion \cite{hologrg,HenSken,dbVV}.  Additional compact directions are represented in the field theory in terms of an infinite tower of composite operators representing the Kaluza-Klein spectrum arising from dimensional reduction of the bulk theory down to the AdS spacetime \cite{adscft2}.  There are other examples in which extra dimensions of space emerge from a strongly coupled field theory -- 
e.g.,
 the (M)atrix model of M-theory \cite{BFSS}.  In all these examples the field theory contains a time, and the emergent gravitational theory inherits its time directly from the field theory.  This leads to an eighth question:  {\it Can time be emergent from the dynamics of a timeless theory?}

The thermodynamic aspects of gravity, and the dualities between gauge theories and gravitational theories, have given rise to a broad idea that spacetime and its metric should generally be thought of as a coarse-grained description of some underlying degrees of freedom which may, or may not, be organized with the proximity and continuity relations associated to smooth spacetime \cite{hologrg,Babeletal,Markonentanglement,Jacobson,Verlinde}.  In these pictures, which are at a preliminary stage of development, the primordial degrees of freedom (the ``atoms of spacetime'', if you will) may be disorganized, in which case there is no smooth spacetime description, or they may cohere to have relations of conditional dependence and entanglement that characterize causal interactions between spatially separated variables.    In the broader quantum gravity community, the notion of causal sets \cite{causalsets} and related approaches \cite{othercausal} view smooth spacetime as an emergent description of relations of conditional dependence of underlying fundamental variables.
This leads to a final question: {\it Are time and space concepts that only become effective in ``phases'' where 
the primordial degrees of freedom self-organize with appropriate relations of conditional dependence and entanglement?}

Below I will describe approaches to some of these questions within string theory.

\section{Towards emergent descriptions of time}

The correspondence between quantum gravity in asymptotically Anti-de Sitter (AdS) spacetimes and conventional conformal field theories (CFTs) is the most concrete setting in which some dimensions of space are ``emergent" from the dynamics of a theory that does not contain these dimensions.  In order to ask whether time can be similarly emergent, it is helpful to first remind ourselves of the prototypical example of this correspondence -- the equivalence between Type IIB string theory compactified on  $S^5$ to AdS$_5$ and the maximally supersymmetric, conformal (32 supercharges) SU(N) Yang-Mills theory defined on the boundary of the AdS$_5$ space \cite{adscft,adscft2}.  Globally, AdS$_5$ is a solid cylinder with a metric 
\be
ds^2 = - (1 + r^2/l^2) \, dt^2 + (1 + r^2/l^2)^{-1} \, dr^2 + r^2 d\Omega_3^2
\Label{ads5met}
\ee
 and the sphere $S^5$ also has a curvature scale $l$.  The boundary of AdS$_5$ ($r \to \infty$) is thus conformal to $S^3 \times R$ with the real line $R$ being time.  Now consider the maximally supersymmetric $SU(N)$ Yang-Mills theory defined on this boundary geometry.  The AdS/CFT correspondence asserts an exact equivalence between string theory (coupling $g_s$ and string length $l_s$) and this Yang-Mills theory with coupling $g^2_{YM} = g_s$ and $N = {1 \over g_s} (l/l_s)^4$.   Thus, the semiclassical limit of spacetime, when  $l \gg l_s$ corresponds to the large $N$ limit of the $SU(N)$ field theory.     

What is the precise dictionary for the correspondence?  First, the symmetries match: AdS$_5 \times S^5$ has an isometry group $SO(4,2) \times SO(6)$ which matches the superconformal group and R-symmetry of the Yang-Mills theory.   String theory in AdS$_5 \times S^5$ has an infinite spectrum of massive fields, some coming from the massive modes of the string, and some from the Kaluza-Klein spectrum of gravity on $S^5$.    A dictionary maps each of these fields onto a specific composite operator of the Yang-Mills theory \cite{adscft2}.   In order to complete the definition of the duality we must specify how observables relate to each other.  This is easiest to state in the field theory limit for the bulk theory.   Consider, for example, a spacetime field $\phi$ that is dual to an operator $O$.  Then
\be
\int_{\phi \to \phi_0}   {\cal D}\phi \, e^{i S(\phi)} \equiv e^{W[\phi_0]} \equiv \int {\cal D}\Phi_{YM} \, e^{i S_{YM} + i\int_{\partial AdS} \phi_0 O} \, 
\ee
where the leftmost equation is the AdS path integral over fields $\phi$ that approach $\phi_0$ at the boundary and $W$ is the generating function of correlation functions of the operator $O$ \cite{adscft2}.  Thus the correlation functions of the Yang-Mills theory are  related to boundary correlation functions of AdS gravity by
\be
\langle O(x_1)... O(x_n) \rangle = {1 \over n^!} \,{\delta^n W[\phi_0] \over \delta \phi_0(x_1) \cdots \phi_0(x_n)} |_{\phi_0 = 0} \, .
\ee
In Lorentzian signature, states of the field theory (e.g. $O | 0 \rangle$) are related to normalizable fluctuations of the dual field $\phi$ which decay near the AdS boundary, while deformations of the field theory by the addition of sources (e.g. $S_{YM} \to S_{YM} + \int \phi_0(x) O(x)$) are related to to non-normalizable changes of boundary condition for bulk fields \cite{BKL}.

How does the radial direction of AdS space ``emerge'' from the dynamics of a quantum field theory on the boundary of space?   There is a beautiful relationship between energy scales in the field theory and radial positions in spacetime \cite{BKLT,hologrg,dbVV,HenSken} wherein, the infrared of the field theory corresponds to the interior of the spacetime, while the ultraviolet corresponds to regions near the boundary.    This can be formalized in the definition of a holographic renormalization group relating the RG flow of the field theory to the radial equations of motion of spacetime \cite{hologrg,HenSken,dbVV}.    In Wilsonian language, integrating out degrees of freedom in the Yang-Mills theory at energy scales above $\Lambda$ corresponds in spacetime to integrating out the skin of AdS space at radii $r \gtrsim  \Lambda \, l^2$ \cite{hologrg}. The M(atrix) model of M-theory provided a similar (though harder to manipulate) ``emergent'' understanding of the ten spatial dimensions of M-theory \cite{BFSS,BGL,PolchMatrix}.   Given these examples, we return to our question -- {\it can we find a setting where time is emergent from a timeless setting?}

To attack this problem one would like to imitate the reasoning that led to the gauge/gravity dualities in a situation where the dual field theory is placed on a Euclidean (and thus timeless) surface at a conformal boundary of spacetime.  Thus we first need a universe which has a non-singular conformal boundary which is Euclidean.  The classic example of such a universe is de Sitter (dS) space, which globally has the metric 
\be
ds^2 = -dt^2 + l^2 \, \cosh^2(t/l)  \, d\Omega_d^2 \, .
\Label{dsmet}
\ee
  Following the example of the gauge/gravity correspondence in AdS space, to have a candidate duality we need to: (a) match symmetries of the dual theories and (b) give a dictionary for calculating correlation functions, (c) match the spectrum of physical states and operators.   To achieve (a) and (b) it suffices to think about the symmetries of the spacetime geometry and to come up with a possible relation between boundary correlators of the spacetime and local correlators of the dual field theory.  For de Sitter space this is made possible by the fact that both de Sitter space (with a positive cosmological constant $\Lambda > 0$) and Euclidean Anti-de Sitter space (with a negative cosmological constant $\Lambda < 0$) are hyperboloids
\be
-x_0^2 + x_1^2 + \cdots + x_D^2 = l^2   ~~~~;~~~  |\Lambda| = {(D-1) (D-2) \over 2l^2 }
\ee
embedded in a flat space with signature (1,D) (see, e.g. \cite{vbds}.  Thus the asymptotic symmetric groups of D-dimensional dS space and of Euclidean AdS space are both equal to the Euclidean conformal group in D dimensions, $SO(D,1$).\footnote{Note however that the quantum physics of de Sitter space cannot be obtained by naive analytic continuation from Euclidean AdS space -- e.g. the two point functions in the vacuum states of the two theories do not continue to each other.}    This led to the suggestion that quantum gravity in de Sitter space is dual a local quantum field theory on  the $D-1$ dimensional sphere  that appears at the conformal boundary  at $t \to \pm \infty$ in (\ref{dsmet}) \cite{stromdscft}.  Imitating the structure of the AdS/CFT correspondence also leads to a proposed formula relating correlators of such a Euclidean conformal field theory (CFT) and scattering amplitudes in de Sitter space \cite{stromdscft, vbds}.

By assuming a dS/CFT correspondence, a number of interesting new features of de Sitter space were discovered, including a way of assigning mass to asymptotically dS universes, and a way of accounting for the entropy of the cosmological horizon seen by inertial observers in such spacetimes \cite{stromdscft,vbds,stromalexdscft}.    Most interestingly, in the conjectural dS/CFT correspondence time evolution was related to the renormalization group flow of the Euclidean field theory, just like the radial flow of the better-understood AdS/CFT correspondence.

While intriguing, the main difficulty with these developments is that there is no concrete realization  in terms of a specific string compactification and thus it is difficult to test the proposed correspondence with precision.   What would it take to {\it derive} or at least argue for a correspondence between a Euclidean field theory and quantum gravity in a time dependent universe from the microscopics of string theory?   The original AdS/CFT correspondence was derived (or at least very well motivated) by considering two different descriptions of the D-brane solitons of string theory \cite{PolchDbrane}.    As an example, recall that the 3-brane at very weak string coupling ($g_s \to 0$) is essentially a 3+1 dimensional sheetlike defect in spacetime and is quantized by considering open strings propagating with endpoints attached to the surface of the defect.    The massless modes of the open strings describe the collective coordinates of the solitons, and, when $N$ 3-branes coincide the massless modes realize a maximally supersymmetric $SU(N)$ gauge theory in 3+1 dimensions (negelecting here the center of mass degrees of freedom of the collection of branes).   On the other hand, when $g_s$ is not infinitesimal the N D-branes are described in terms of closed strings propagating in a background with a metric
\be
ds^2 = (1 + Q/r^4)r^{-1/2} (-dt^2 + dx_1^2 + \cdots dx_3^2) +  (1 + Q/r^4)r^{1/2} (dr^2 + r^2 d\Omega_5^2)
\Label{3-brane}
\ee
where $Q \propto g_s N l_s^4$ is the charge of the D-branes.
The open string and closed string descriptions are equivalent by the open-closed dualities of string theory.  The key insight that led to the AdS/CFT correspondence was that there is a decoupling limit in which the low-energy theory of the open strings (the $SU(N)$ Yang-Mills propagating on the brane) is by itself equivalent to the theory of closed strings propagating in the ``throat'' or near-horizon region of the 3-brane spacetime \cite{adscft}.  This near-horizon region is isolated essentially by dropping the 1 in the overall factors in (\ref{3-brane}) and leads to an AdS$_5 \times S^5$ geometry:
\be
ds^2 = {l^2 \over r^2} (-dt^2 + d\vec{x}^2) + {r^2 \over l^2} dr^2 + l^2 d\Omega_5^2
\Label{ads5again}
\ee
where $l^2 = \sqrt{Q} \sim (g_s N) l_s$ is the curvature scale of the geometry, and the first two terms give AdS$_5$ space (in different coordinates than (\ref{ads5met})).   Thus a four dimensional field theory is argued to be equivalent to quantum gravity in the background (\ref{ads5again}) with the radial direction and the 5-sphere emerging out of the field theory dynamics as described earlier.

 Can some similar argument relate a Euclidean field theory to a time-dependent, perhaps de Sitter-like, geometry?  To achieve this we first need some kind of Euclidean D-brane -- i.e. a ``defect'' that is localized in time and that is quantized by open strings attached to such a spacelike surface.    Natural candidates for such ``SD-branes" were proposed in \cite{spacelikebranes} and \cite{rollingtachyon} where the dynamical condensation of the tachyon on a unstable D-brane was considered.  As such D-branes decay, they disappear entirely, leaving behind a background of closed string radiation propagating in a time dependent geometry.    In fact, one of the key motivations in \cite{spacelikebranes} for studying such objects was the hope that, in analogy, with the derivation of the AdS/CFT correspondence with an emergent spatial direction, SD-branes would lead to an emergent time direction.   This program has not yet been taken to completion as there have been many challenges -- e.g. the spacetime geometry for N coincident decaying branes has not been brought under sufficient control (although see \cite{nonsingsbrane}) as to permit a recap of the AdS/CFT logic relating a ``near-brane'' geometry to the dynamics of open strings localized on the brane.  However, there are some intriguing hints.

For example, the authors of \cite{naqvilarsen} and \cite{matrixtime1} considered  a class of unstable branes where the soliton is understood to be present at the ``beginning of time'' and then decays away, so that the open strings that quantize the branes are essentially localized to the initial instant of the time.  These papers showed that perturbative open string calculations of the scattering amplitudes from a single decaying brane can be reduced to a sum of correlators computed in a grand canonical ensemble of unitary random matrix models with time setting the fugacity.  Interestingly, the later the time, the larger the fugacity -- thus, as the brane decays away leaving only closed strings behind, the system is dominated by the dynamics of matrices of larger and larger rank.    This is intriguing because in the AdS/CFT correspondence extra dimensions of spacetime emerge precisely from the dynamics of large unitary matrices.   Indeed, going back to the original arguments of 't Hooft concerning large N gauge theory, it has been appreciated that large unitary matrices have a relation to the dynamics of closed strings, which have since been understood to be associated to an emergent spacetime.  This suggests that in the theory of decaying branes, time and the geometry at late times might be ``emerging'' from the dynamics of the large matrices appearing in the open string description.   Subsequent work \cite{thermotime} showed that this analogy can be pushed further -- the free energy of the ensemble of matrices decreases towards later times, defining a thermodynamic arrow of time, and the time evolution equations of spacetime get mapped to differential equations relating expectation values of observables at different points of thermal equilibrium.

Where are the large unitary matrices coming from?  In another intriguing connection, it was shown that unstable branes can be equivalently understood in terms of arrays of branes localized in complexified time \cite{gaiotto}.\footnote{As discussed in the introduction, this raises again the issue of whether we should regard conventional real spacetime as simply a convenient way of summarizing analytic information about an underlying complexified structure.}   Perhaps it is possible to use this insight to implement an AdS/CFT like decoupling limit that relates the theory of the lightest modes on these D-branes to the early-time (and hence near-brane) geometry of the unstable D-brane \cite{usongoing}.   Success with this program would give the first concrete example of a model where time emerges from the dynamics of a timeless theory.

\section{Towards timeless states of the universe}

It is interesting to ask whether it is possible to construct and understand ``phases'' of quantum  gravity where time ceases to exist.   One interesting setting where this question arises is in the ``bubble of nothing'' spacetimes.   Consider a flat  five dimensional universe with one circular spatial dimension ($R^{1,3} \times S^1$) where supersymmetry is broken.  Following Witten \cite{wittennothing}, it can be shown that such a universe is unstable to tunneling to ``nothing''.  More precisely, there is an instanton (constructed by double Wick rotation of the fived-dimensional Schwarzschild solution) which mediates tunneling from $R^{1,3} \times S^1$ to a configuration where a hole (the ``bubble of nothing'') appears in the classical spacetime.  This configuration is smooth -- the $S^1$ shrinks in size as the hole is approached in such a way that the entire five dimensional geometry is non-singular.  The hole then expands to eventually consume all of spacetime.   Evidently conventional spacetime does not exist inside the bubble of nothing.  But is it really ``nothing'' or is it some incoherent phase which lacks a spacetime interpretation?

A way of attacking this question is to find a bubble of nothing instability Anti-de Sitter space which then has an interpretation in a dual quantum field theory.  By  understanding the corresponding field theoretic instability we might get an insight into the nothing state.   Such an embedding was carried out in \cite{adsnothing} which considered instabilities of a ``topological black hole'' constructed by identifying global AdS$_5$ space under a boost  \cite{topblack}.   This geometry has a horizon at $r=r_+$, outside which the metric can be written as
\be
ds^2 = {l^2 \over r^2 - r_+^2} \, dr^2 + (r^2 - r_+^2) \left[-dt^2 + {l^2 \over r_+^2} \cosh^2\left({r_+ t \over l}\right) \, d\Omega_2 \right] + r^2 \, d\chi^2
\Label{bubblefalse}
\ee
where $l$ is the AdS curvature scale, $d\Omega_2^2$ is the line element on the 2-sphere and $\chi$ is a circle with $\chi \sim \chi + 2\pi$.  Note that the radial sections, and thus the conformal boundary at $r \to \infty$, of this geometry are de Sitter space times a circle.   Thus, the topological black hole should be dual to SU(N) Yang-Mills theory on three-dimensional de Sitter space times circle of fixed size.     The authors of \cite{adsnothing} proceeded to show that when $r_+ < l/2\sqrt{2}$, the topological black hole (\ref{bubblefalse}) is unstable and tunnels via an instanton to a bubble of nothing in AdS space.  Like Witten's bubble of nothing in flat space \cite{wittennothing}, this hole in AdS space expands outward and eventually consumes the classical geometry.   However, from the perspective of the dual field theory we are simply dealing with a conventional quantum field theory in a curved spacetime (de Sitter times a circle).  So what is ``nothing'' from this dual point of view?  It was argued in \cite{adsnothing,moshenothing} that the true vacuum of the Yang-Mills theory in this background corresponds to a condensate of a winding mode on the circle.  Thus the decay to non-geometrical ``nothing'' in spacetime simply corresponded to phase transition to a phase of the dual field theory where the winding mode condenses, disrupting the order that allows the bulk AdS geometry to emerge as a description of the field theory.  A related picture emerged in \cite{evagary}, whose authors suggested that a condensate arising from the dynamics of string-theoretic unstable modes results in a non-geometric final state replacing the singularity of certain black holes.

The idea that smooth spacetime can sometimes be seen as an ordered phase of underlying degrees of freedom raises the question whether singularities where time begins or ends (e.g. inside a Schwarzschild black hole or at the Big Bang) can be understood as demarcating a boundary between between ordered and disordered phases in some way.  This is especially interesting in the context of the question ``What came `before' the Big Bang?'', and cosmological scenarios like ekpyrosis \cite{ekpyrosis} where the universe bounces through a singularity.   The methods of perturbative string theory are unlikely to be helpful because perturbation theory breaks down near a spacetime singularity.   Thus, once again, much of the thinking about such singularities has occurred in the context of gauge/gravity dualities, either in the AdS/CFT correspondence or the M(atrix) model of M-theory.   

In the AdS/CFT correspondence it is easy to construct Schwarzschild black holes, which are dual to a field theory at finite temperature.   Within this field theory one can ask how physics behind the horizon and near the singularity is represented.   This question is most conveniently asked in terms of the 2-point correlation function of  operators ${\cal O}$ with a large conformal dimension $\Delta$.   In the AdS gravity, this correlator can be expressed, in the semiclassical limit, in terms of the lengths ${\cal L}(\vec{x},t)$ of bulk geodesics that connect the two points on the spacetime boundary where the operators are placed: $\langle {\cal O}(\vec{x},t) {\cal O}(0,0) \rangle \sim exp[-\Delta {\cal L}(\vec{x}, t)$ \cite{simonvijaygeodesic}.  Using this technique the authors of \cite{behindthehorizon} found evidence that singularities of the classical geometry were encoded in the analytic structure of the pair correlation function of the dual large $N$ field theory in the complexified coordinate plane, again suggesting a key role for an underlying complexified geometry.    Unfortunately, it has not been possible in this approach to understand how the Schwarzschild singularity is resolved in the underlying quantum theory since the semiclassical calculations assume the existence of the standard Schwarzschild geometry.   Understanding the resolution of such singularities remains a key question in string theory.


%

\section{Conclusion}

As is evident from the discussion above, we have many more questions about time than answers.   One of the challenges is that many of the deepest questions require a non-perturbative formulation of string theory and we only really have that in the context of the AdS/CFT correspondence.  This is why the concrete approaches described above mostly used this tool.    I have selectively reviewed some of the pertinent developments, focussing on the approaches that I feel to be most provocative.  But there are many other leads to follow. To name one, there has been extensive development of space-time orbifolds in string theory as toy models of time-dependent spacetimes where questions of singularity resolution can be studied \cite{spacetimeorbifold}.    Likewise, there has been extensive study of a class of null (lightlike) singularities in string theory which admit a dual description in terms of the M(atrix) model of M-theory \cite{Benandco}.   Within the broader quantum gravity community outside string theory there has also been considerable thinking about time - e.g., models where space time is emergent are discussed in \cite{causalsets,othercausal}.  Traditionally, in the study of quantum gravity the ``problem of time'' \cite{problemoftime} arises because the Schr\"{o}dinger equation ($\hat{H} | \psi \rangle = i \partial_t | \psi \rangle$), when promoted to the diffeomorphism invariant context of gravity, becomes the Wheeler-de Witt equation, $\hat{H}|\psi \rangle = 0$, which simply provides a constraint on the possible states of the universe and says nothing about time evolution.    This is sometimes interpreted as saying that saying that in a quantum diffeomorphism-invariant universe time is meaningless (for discussions of this idea and references see \cite{problemoftime}.)

One can hope that the next forty years of string theory will see progress on the nine questions listed in the introduction.

\section*{Acknowledgements}
I have enjoyed discussions with  Bartek Czech, Djordje Minic, and Moshe Rozali about this material.   I also thank the Lovers \& Madmen cafe in Philadelphia, the Pitkin County Library in Aspen and the Aspen Center for Physics for hospitality during completion of the paper.


\begin{thebibliography}{..}

\Bibitem{HawkingCTC}
  S.~W.~Hawking,
  ``The Chronology protection conjecture,''
  Phys.\ Rev.\  {\bf D46}, 603-611 (1992);~~~
  S.~W.~Hawking,
  ``Quantum coherence and closed timelike curves,''
  Phys.\ Rev.\  {\bf D52}, 5681-5686 (1995).
  [gr-qc/9502017].


\bibitem{ArrowOfTime}
  S.~M.~Carroll, J.~Chen,
  ``Spontaneous inflation and the origin of the arrow of time,''
  [hep-th/0410270];!!!
  H.~Price,
  `Cosmology, time's arrow, and that old double standard,''
  [gr-qc/9310022];~~~
  A.~Albrecht,
  ``Cosmic inflation and the arrow of time,''
  In *Barrow, J.D. (ed.) et al.: Science and ultimate reality* 363-401.
  [astro-ph/0210527].


\Bibitem{twistor} 
  E.~Witten,
  ``Perturbative gauge theory as a string theory in twistor space,''
  Commun.\ Math.\ Phys.\  {\bf 252}, 189-258 (2004).
  [hep-th/0312171].


\Bibitem{twotimes}
  I.~Bars,
  ``Survey of two time physics,''
  Class.\ Quant.\ Grav.\  {\bf 18}, 3113-3130 (2001).
  [hep-th/0008164].
  
  \bibitem{hull}
    C.~M.~Hull,
  ``Timelike T duality, de Sitter space, large N gauge theories and topological field theory,''
  JHEP {\bf 9807}, 021 (1998).
  [hep-th/9806146];~~~
  C.~M.~Hull, R.~R.~Khuri,
  ``Branes, times and dualities,''
  Nucl.\ Phys.\  {\bf B536}, 219-244 (1998).
  [hep-th/9808069];~~~
  C.~M.~Hull,
  ``Duality and the signature of space-time,''
  JHEP {\bf 9811}, 017 (1998).
  [hep-th/9807127].


\Bibitem{yangmillsamplitudes}
  R.~Britto, F.~Cachazo, B.~Feng, E.~Witten,
  ``Direct proof of tree-level recursion relation in Yang-Mills theory,''
  Phys.\ Rev.\ Lett.\  {\bf 94}, 181602 (2005).
  [hep-th/0501052].



\Bibitem{behindthehorizon} 
  P.~Kraus, H.~Ooguri, S.~Shenker,
  ``Inside the horizon with AdS / CFT,''
  Phys.\ Rev.\  {\bf D67}, 124022 (2003).
  [hep-th/0212277];~~~
  L.~Fidkowski, V.~Hubeny, M.~Kleban, S.~Shenker,
  ``The Black hole singularity in AdS / CFT,''
  JHEP {\bf 0402}, 014 (2004).
  [hep-th/0306170];~~~
    T.~S.~Levi, S.~F.~Ross,
  ``Holography beyond the horizon and cosmic censorship,''
  Phys.\ Rev.\  {\bf D68}, 044005 (2003).
  [hep-th/0304150];~~~
    V.~Balasubramanian, T.~S.~Levi,
  ``Beyond the veil: Inner horizon instability and holography,''
  Phys.\ Rev.\  {\bf D70}, 106005 (2004).
  [hep-th/0405048].

\Bibitem{fuzzballsolns}
  I.~Bena, N.~Bobev, S.~Giusto, C.~Ruef, N.~P.~Warner,
  ``An Infinite-Dimensional Family of Black-Hole Microstate Geometries,''
  JHEP {\bf 1103}, 022 (2011).
  [arXiv:1006.3497 [hep-th]];~~~
  I.~Bena, N.~P.~Warner,
  ``Bubbling supertubes and foaming black holes,''
  Phys.\ Rev.\  {\bf D74}, 066001 (2006).
  [hep-th/0505166];~~~
  P.~Berglund, E.~G.~Gimon, T.~S.~Levi,
  ``Supergravity microstates for BPS black holes and black rings,''
  JHEP {\bf 0606}, 007 (2006).
  [hep-th/0505167];~~~
  E.~G.~Gimon, F.~Larsen, J.~Simon,
  ``Black holes in Supergravity: The Non-BPS branch,''
  JHEP {\bf 0801}, 040 (2008).
  [arXiv:0710.4967 [hep-th]];~~~ 
    O.~Lunin, S.~D.~Mathur,
  ``Metric of the multiply wound rotating string,''
  Nucl.\ Phys.\  {\bf B610}, 49-76 (2001).
  [hep-th/0105136].



\Bibitem{Dbranes}
See the review and references in  A.~W.~Peet,
  ``TASI lectures on black holes in string theory,''
  [hep-th/0008241].


\Bibitem{topsing}
  For example, see C.~V.~Johnson, A.~W.~Peet, J.~Polchinski,
  ``Gauge theory and the excision of repulson singularities,''
  Phys.\ Rev.\  {\bf D61}, 086001 (2000).
  [hep-th/9911161];~~~    H.~Lin, O.~Lunin, J.~M.~Maldacena,
  ``Bubbling AdS space and 1/2 BPS geometries,''
  JHEP {\bf 0410}, 025 (2004).
  [hep-th/0409174].



\Bibitem{conifold}
  A.~Strominger,
  ``Massless black holes and conifolds in string theory,''
  Nucl.\ Phys.\  {\bf B451}, 96-108 (1995).
  [hep-th/9504090].


\Bibitem{BHthermo}
  J.~M.~Bardeen, B.~Carter, S.~W.~Hawking,
  ;``The Four laws of black hole mechanics,''
  Commun.\ Math.\ Phys.\  {\bf 31}, 161-170 (1973);~~~
    J.~D.~Bekenstein,
  ``Black holes and entropy,''
  Phys.\ Rev.\  {\bf D7}, 2333-2346 (1973);~~~
    S.~W.~Hawking,
  ``Particle Creation by Black Holes,''
  Commun.\ Math.\ Phys.\  {\bf 43}, 199-220 (1975).
  




\Bibitem{bhentropy}
  A.~Strominger, C.~Vafa,
  ``Microscopic origin of the Bekenstein-Hawking entropy,''
  Phys.\ Lett.\  {\bf B379}, 99-104 (1996).
  [hep-th/9601029].




\Bibitem{infoloss}
  S.~W.~Hawking,
  ``Breakdown of Predictability in Gravitational Collapse,''
  Phys.\ Rev.\  {\bf D14}, 2460-2473 (1976).
  

\Bibitem{inforecovery} 
  V.~Balasubramanian, B.~Czech,
  ``Quantitative approaches to information recovery from black holes,''
  Class.\ Quan. Grav. \ {\bf 28}, 163001  (2011).  
  [arXiv:1102.3566 [hep-th]].
  
  

  
  
  \Bibitem{fuzzball}
    S.~D.~Mathur,
  ``Fuzzballs and the information paradox: A Summary and conjectures,''
  [arXiv:0810.4525 [hep-th]];~~~
    S.~D.~Mathur,
  ``The Fuzzball proposal for black holes: An Elementary review,''
  Fortsch.\ Phys.\  {\bf 53}, 793-827 (2005).
  [hep-th/0502050].

\Bibitem{noinside}
  V.~Balasubramanian, D.~Marolf, M.~Rozali,
  ``Information Recovery From Black Holes,''
  Gen.\ Rel.\ Grav.\  {\bf 38}, 1529-1536 (2006).
  [hep-th/0604045];~~~
  V.~Balasubramanian, B.~Czech, V.~E.~Hubeny, K.~Larjo, M.~Rangamani, J.~Simon,
  ``Typicality versus thermality: An Analytic distinction,''
  Gen.\ Rel.\ Grav.\  {\bf 40}, 1863-1890 (2008).
  [hep-th/0701122];~~~
    G.~Horowitz, A.~Lawrence, E.~Silverstein,
  ``Insightful D-branes,''
  JHEP {\bf 0907}, 057 (2009).
  [arXiv:0904.3922 [hep-th]].



  \Bibitem{albionevagary}
  G.~Horowitz, A.~Lawrence, E.~Silverstein,
  ``Insightful D-branes,''
  JHEP {\bf 0907}, 057 (2009).
  [arXiv:0904.3922 [hep-th]].

 
 
 \Bibitem{Babeletal}
  V.~Balasubramanian, J.~de Boer, V.~Jejjala, J.~Simon,
  ``The Library of Babel: On the origin of gravitational thermodynamics,''
  JHEP {\bf 0512}, 006 (2005).
  [hep-th/0508023];~~~
    V.~Balasubramanian, B.~Czech, K.~Larjo, D.~Marolf, J.~Simon,
  ``Quantum geometry and gravitational entropy,''
  JHEP {\bf 0712}, 067 (2007).
  [arXiv:0705.4431 [hep-th]]


\Bibitem{complementarity}
  L.~Susskind,
  ``String theory and the principles of black hole complementarity,''
  Phys.\ Rev.\ Lett.\  {\bf 71}, 2367-2368 (1993).
  [hep-th/9307168].


\Bibitem{adscft}
  J.~M.~Maldacena,
  ``The Large N limit of superconformal field theories and supergravity,''
  Adv.\ Theor.\ Math.\ Phys.\  {\bf 2}, 231-252 (1998).
  [hep-th/9711200].


\Bibitem{adscft2} 
  S.~S.~Gubser, I.~R.~Klebanov, A.~M.~Polyakov,
  ``Gauge theory correlators from noncritical string theory,''
  Phys.\ Lett.\  {\bf B428}, 105-114 (1998).
  [hep-th/9802109];~~~~
  E.~Witten,
  ``Anti-de Sitter space and holography,''
  Adv.\ Theor.\ Math.\ Phys.\  {\bf 2}, 253-291 (1998).
  [hep-th/9802150].

\Bibitem{hologrg}
  V.~Balasubramanian, P.~Kraus,
  ``Space-time and the holographic renormalization group,''
  Phys.\ Rev.\ Lett.\  {\bf 83}, 3605-3608 (1999).
  [hep-th/9903190].

\Bibitem{HenSken} 
  M.~Henningson, K.~Skenderis,
  ``The Holographic Weyl anomaly,''
  JHEP {\bf 9807}, 023 (1998).
  [hep-th/9806087];~~~
  K.~Skenderis,
  ``Lecture notes on holographic renormalization,''
  Class.\ Quant.\ Grav.\  {\bf 19}, 5849-5876 (2002).
  [hep-th/0209067].


\Bibitem{dbVV}  J.~de Boer, E.~P.~Verlinde, H.~L.~Verlinde,
  ``On the holographic renormalization group,''
  JHEP {\bf 0008}, 003 (2000).
  [hep-th/9912012].






\Bibitem{BFSS} 
  T.~Banks, W.~Fischler, S.~H.~Shenker, L.~Susskind,
  ``M theory as a matrix model: A Conjecture,''
  Phys.\ Rev.\  {\bf D55}, 5112-5128 (1997).
  [hep-th/9610043].





\Bibitem{Markonentanglement} 
  M.~Van Raamsdonk,
  ``Building up spacetime with quantum entanglement,''
  Gen.\ Rel.\ Grav.\  {\bf 42}, 2323-2329 (2010).
  [arXiv:1005.3035 [hep-th]];~~~
    M.~Van Raamsdonk,
  ``Comments on quantum gravity and entanglement,''  
  [arXiv:0907.2939 [hep-th]].


\Bibitem{Jacobson} 
  T.~Jacobson,
  ``Thermodynamics of space-time: The Einstein equation of state,''
  Phys.\ Rev.\ Lett.\  {\bf 75}, 1260-1263 (1995).
  [gr-qc/9504004].


\Bibitem{Verlinde} 
  E.~P.~Verlinde,
  ``On the Origin of Gravity and the Laws of Newton,''
  JHEP {\bf 1104}, 029 (2011).
  [arXiv:1001.0785 [hep-th]].

\Bibitem{causalsets}
  L.~Bombelli, J.~Lee, D.~Meyer, R.~Sorkin,
  ``Space-Time as a Causal Set,''
  Phys.\ Rev.\ Lett.\  {\bf 59}, 521-524 (1987).


\Bibitem{othercausal}
  T.~Konopka, F.~Markopoulou, L.~Smolin,
  ``Quantum Graphity,''
  [hep-th/0611197];~~~
    T.~Konopka, F.~Markopoulou, S.~Severini,
  ``Quantum Graphity: A Model of emergent locality,''
  Phys.\ Rev.\  {\bf D77}, 104029 (2008).
  [arXiv:0801.0861 [hep-th]].
  
  
  



\Bibitem{BKL}   V.~Balasubramanian, P.~Kraus, A.~E.~Lawrence,
  ``Bulk versus boundary dynamics in anti-de Sitter space-time,''
  Phys.\ Rev.\  {\bf D59}, 046003 (1999).
  [hep-th/9805171].


\Bibitem{BKLT}   V.~Balasubramanian, P.~Kraus, A.~E.~Lawrence, S.~P.~Trivedi,
  ``Holographic probes of anti-de Sitter space-times,''
  Phys.\ Rev.\  {\bf D59}, 104021 (1999).
  [hep-th/9808017].




\Bibitem{AdSstress}   V.~Balasubramanian, P.~Kraus,
  ``A Stress tensor for Anti-de Sitter gravity,''
  Commun.\ Math.\ Phys.\  {\bf 208}, 413-428 (1999).
  [hep-th/9902121].





\Bibitem{BGL}   V.~Balasubramanian, R.~Gopakumar, F.~Larsen,
  ``Gauge theory, geometry and the large N limit,''
  Nucl.\ Phys.\  {\bf B526}, 415-431 (1998).
  [hep-th/9712077].


\Bibitem{PolchMatrix}    J.~Polchinski,
  ``M theory and the light cone,''
  Prog.\ Theor.\ Phys.\ Suppl.\  {\bf 134}, 158-170 (1999).
  [hep-th/9903165].



\Bibitem{stromdscft} 
  A.~Strominger,
  ``The dS / CFT correspondence,''
  JHEP {\bf 0110}, 034 (2001).
  [hep-th/0106113].


\Bibitem{vbds} 
  V.~Balasubramanian, J.~de Boer, D.~Minic,
  ``Mass, entropy and holography in asymptotically de Sitter spaces,''
  Phys.\ Rev.\  {\bf D65}, 123508 (2002).
  [hep-th/0110108];~~~
    V.~Balasubramanian, J.~de Boer, D.~Minic,
  ``Notes on de Sitter space and holography,''
  Class.\ Quant.\ Grav.\  {\bf 19}, 5655-5700 (2002).
  [hep-th/0207245]

\Bibitem{stromalexdscft} 
  R.~Bousso, A.~Maloney, A.~Strominger,
  ``Conformal vacua and entropy in de Sitter space,''
  Phys.\ Rev.\  {\bf D65}, 104039 (2002).
  [hep-th/0112218].



\Bibitem{PolchDbrane}  
  J.~Polchinski,
  ``Dirichlet Branes and Ramond-Ramond charges,''
  Phys.\ Rev.\ Lett.\  {\bf 75}, 4724-4727 (1995).
  [hep-th/9510017].


\Bibitem{spacelikebranes} 
  M.~Gutperle, A.~Strominger,
  ``Space - like branes,''
  JHEP {\bf 0204}, 018 (2002).
  [hep-th/0202210].


\Bibitem{nonsingsbrane}
  G.~Jones, A.~Maloney, A.~Strominger,
  ``Nonsingular solutions for S-branes,''
  Phys.\ Rev.\  {\bf D69}, 126008 (2004).
  [hep-th/0403050].


\Bibitem{rollingtachyon} 
  A.~Sen,
  ``Rolling tachyon,''
  JHEP {\bf 0204}, 048 (2002).
  [hep-th/0203211].



\Bibitem{naqvilarsen}
  F.~Larsen, A.~Naqvi, S.~Terashima,
  ``Rolling tachyons and decaying branes,''
  JHEP {\bf 0302}, 039 (2003).
  [hep-th/0212248].


\Bibitem{lambert} 
  N.~D.~Lambert, H.~Liu, J.~M.~Maldacena,
  ``Closed strings from decaying D-branes,''
  JHEP {\bf 0703}, 014 (2007).
  [hep-th/0303139].


\Bibitem{gaiotto} 
  D.~Gaiotto, N.~Itzhaki, L.~Rastelli,
  ``Closed strings as imaginary D-branes,''
  Nucl.\ Phys.\  {\bf B688}, 70-100 (2004).
  [hep-th/0304192].


\Bibitem{matrixtime1} 
  V.~Balasubramanian, E.~Keski-Vakkuri, P.~Kraus, A.~Naqvi,
  ``String scattering from decaying branes,''
  Commun.\ Math.\ Phys.\  {\bf 257}, 363-394 (2005).
  [hep-th/0404039].

\Bibitem{thermotime} 
  V.~Balasubramanian, N.~Jokela, E.~Keski-Vakkuri, J.~Majumder,
  ``A Thermodynamic interpretation of time for rolling tachyons,''
  Phys.\ Rev.\  {\bf D75}, 063515 (2007).
  [hep-th/0612090].


\Bibitem{usongoing} B.~Czech, M.~Rozali and V.~Balasubramanian, In progress.


\Bibitem{wittennothing} 
  E.~Witten,
  ``Instability of the Kaluza-Klein Vacuum,''
  Nucl.\ Phys.\  {\bf B195}, 481 (1982).


\Bibitem{adsnothing}
  V.~Balasubramanian, S.~F.~Ross,
  ``The Dual of nothing,''
  Phys.\ Rev.\  {\bf D66}, 086002 (2002).
  [hep-th/0205290];~~~
  V.~Balasubramanian, K.~Larjo, J.~Simon,
  ``Much ado about nothing,''
  Class.\ Quant.\ Grav.\  {\bf 22}, 4149-4170 (2005).
  [hep-th/0502111].
  
\Bibitem{topblack}
  M.~Banados, A.~Gomberoff, C.~Martinez,
  ``Anti-de Sitter space and black holes,''
  Class.\ Quant.\ Grav.\  {\bf 15}, 3575-3598 (1998).
  [hep-th/9805087].
  
  
 \Bibitem{moshenothing}
   J.~He, M.~Rozali,
  ``On bubbles of nothing in AdS/CFT,''
  JHEP {\bf 0709}, 089 (2007).
  [hep-th/0703220 [HEP-TH]].

 
 \Bibitem{evagary}
   G.~T.~Horowitz, E.~Silverstein,
  ``The Inside story: Quasilocal tachyons and black holes,''
  Phys.\ Rev.\  {\bf D73}, 064016 (2006).
  [hep-th/0601032
 
\Bibitem{ekpyrosis}
  J.~Khoury, B.~A.~Ovrut, P.~J.~Steinhardt, N.~Turok,
  ``The Ekpyrotic universe: Colliding branes and the origin of the hot big bang,''
  Phys.\ Rev.\  {\bf D64}, 123522 (2001).
  [hep-th/0103239].


\Bibitem{simonvijaygeodesic}    V.~Balasubramanian, S.~F.~Ross,
  ``Holographic particle detection,''
  Phys.\ Rev.\  {\bf D61}, 044007 (2000).
  [hep-th/9906226].



\Bibitem{spacetimeorbifold}
  H.~Liu, G.~W.~Moore and N.~Seiberg,
  ``Strings in a time dependent orbifold,''
  JHEP {\bf 0206} (2002) 045
  [arXiv:hep-th/0204168];~~~
    J.~Simon,
  ``The Geometry of null rotation identifications,''
  JHEP {\bf 0206} (2002) 001
  [arXiv:hep-th/0203201].
  

\Bibitem{Benandco} 
 B.~Craps, S.~Sethi, E.~P.~Verlinde,
  ``A Matrix big bang,''
  JHEP {\bf 0510}, 005 (2005).
  [hep-th/0506180];~~~  B.~Craps, O.~Evnin,
  ``Light-like Big Bang singularities in string and matrix theories,''
  [arXiv:1103.5911 [hep-th]], to appear in Class. Quant. Grav.

\Bibitem{problemoftime} 
    S.~M.~Carroll,
  ``What if Time Really Exists?,''
  [arXiv:0811.3772 [gr-qc]];~~~
    F.~Markopoulou,
  ``Space does not exist, so time can,''
  [arXiv:0909.1861 [gr-qc]];~~~
 C.~Rovelli,
  ``'Forget time',''
  [arXiv:0903.3832 [gr-qc]].


\end{thebibliography}
\end{document}